\documentclass{WileyMSP-template}
\usepackage[breaklinks=true,unicode=true,urlcolor = blue,colorlinks = true,citecolor = blue,linkcolor = blue]{hyperref}

\begin{document}

\pagestyle{fancy}
\rhead{\includegraphics[width=2.5cm]{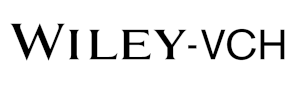}}

\title{Unraveling the Complexity of the Dzyaloshinskii--Moriya Interaction in Layered Magnets: The Full Magnitude and Chirality Control}

\maketitle


\author{Khalil Zakeri\href{mailto:khalil.zakeri@kit.edu}{*}}
\author{Albrecht von Faber}
\author{Sergiy Mankovsky}
\author{Hubert Ebert}



\begin{affiliations}
Khalil Zakeri, Albrecht von Faber\\
Heisenberg Spin-dynamics Group, Physikalisches Institut, Karlsruhe Institute of Technology, Wolfgang-Gaede-Strasse 1, D-76131 Karlsruhe, Germany\\
Email Address: {\href{mailto:khalil.zakeri@kit.edu}{khalil.zakeri@kit.edu}}

Sergiy Mankovsky, Hubert Ebert\\
Department of Chemistry and Physical Chemistry, LMU Munich, Butenandtstrasse 11, D-81377 Munich, Germany

\end{affiliations}



\normalfont

\begin{abstract}

{\sffamily \textbf{Chirality is one of the inherent characteristics of some objects in nature. In magnetism, chiral magnetic textures can be formed in systems with broken inversion symmetry and due to an antisymmetric magnetic interaction, known as Dzyaloshinskii–Moriya interaction (DMI). Here, aiming for a fundamental understanding of this chiral interaction on the atomic scale, several synthetic layered structures composed of alternating atomic layers of 3d ferromagnetic metals epitaxially grown on the Ir(001) surface are designed. It is demonstrated both experimentally and theoretically that the atomistic DMI depends critically not only on the orbital occupancy of the interface magnetic layer but also on the sequence of the atomic layers. It is shown that even large atomistic DMI values can result in a small effective DMI, and conversely. Furthermore, the dependence of the effective DMI on the number of atomic layers deviates from a simple scaling law. These observations are attributed to the complexity of the electronic structure and the contributions of different orbitals to the hybridization and DMI. The results are anticipated to provide guidelines for achieving full control over both the chirality and the magnitude of the atomistic DMI in layered materials.}}

\end{abstract}


\section{Introduction}

Chirality is a fundamental symmetry phenomenon in nature. The DNA, amino acids, nucleic acids and many carbohydrates are chiral objects. Chirality plays a crucial role in the interaction between enzymes and their substrates, a process which is essential for many chemical reactions on which the life is based \cite{Naaman2011}.
Chiral objects may also be formed in magnetic systems possessing a chiral magnetic interaction. Such an antisymmetric interaction,  known as Dzyaloshinskii--Moriya interaction (DMI), is present in spin systems with broken inversion symmetry and a large spin-orbit coupling (SOC) \cite{Dzyaloshinsky1958}. In a spin Hamiltonian representation DMI is given by $\mathcal{H_{\mathrm{DMI}}}=\sum_{i\neq j}\vec{D}_{ij}\cdot \vec{S}_i \times \vec{S}_j$, where the chirality of the interaction between spins $\vec{S}_i$ and $\vec{S}_j$ is determined by the direction of the DM vectors $\vec{D}_{ij}$ \cite{Moriya1960}.  Similar to the symmetric Heisenberg exchange interaction, DMI is also an inherently quantum mechanical effect \cite{Dzyaloshinsky1958, Moriya1960}. The most important characteristic of this interaction is that it is  chiral and leads to a twist in the spin system \cite{Moriya1960}. Hence, a detailed knowledge of this interaction would provide an access to the chirality and size of topological spin textures formed in chiral magnetic materials \cite{Grigoriev2013,Shibata2013,Siegfried2015,Beutier2017} and enable the prediction of their dynamics under external stimuli \cite{Abdizadeh2020}.

When monolayers of $3d$ ferromagnetic metals such as Fe, Co and Ni are grown on substrates with a large SOC, one expects a sufficiently large DMI, leading to the formation of chiral spin textures \cite{Fert1980,Crepieux1998,Roessler2006,Bode2007,Vedmedenko2007,Heinze2011,Fert2017}. Similar to many other magnetic parameters, it has been observed that DMI is also very sensitive to the changes in the electronic structure of the system induced by external perturbations. Recently, a strong DMI induced by chemisorption of oxygen on the magnetic layers has been reported \cite{Chen2020}. In a similar way, it has been demonstrated that other mechanisms, such as variations in chemical composition  \cite{Kim2019,Zhang2022,Liang2022} or introducing inhomogeneous strain effects \cite{Davydenko2019,Zhang2021}, can also be used to tune DMI in both magnetic thin films and bulk magnets. These mechanisms provide alternative ways to artificially engineer DMI, extending beyond the capabilities of standard approaches based on intrinsic mechanisms.
	
Using theoretical calculations, Yang, \textit{et al.} have explored the influence of layer stacking on the DMI, with a specific focus on nonmagnetic layers like Pt and Ir. \cite{Yang2015,Yang2018}. The calculations have been performed for the (111) surface orientation of the layers indicating a possible tuning of DMI by changing the number and combinations of the magnetic and nonmagnetic layers. Nonetheless, the investigation of the impact of the ferromagnetic layer stacking on the atomistic DMI within the ultrathin regime remains unexplored, in particular from an experimental perspective.
	
The atomistic DMI in planar magnets displays a complex pattern. Recent studies have demonstrated that $\vec{D}_{ij}s$  in planar systems can exhibit a chirality inversion, depending on the distance between interacting spins \cite{Zakeri2023a}. The situation is anticipated to become increasingly intricate as the number of atomic layers increases \cite{Yang2015,Tsurkan2020}. Hence, layered magnetic systems with different atomic layers would provide a platform to control the complex pattern of $\vec{D}_{ij}$ and, consequently, the effective DMI throughout a combination of the symmetry modification and additive/subtractive effects of $\vec{D}_{ij}$s. In this respect it is of prime importance to unravel the impact of the number of the $d$-state electrons on $\vec{D}_{ij}$s. Such a systematic experimental investigation of epitaxial atomic layers with well-defined structures, as model systems, is still missing. In the next step one requires to experimentally investigate the impact of the number and sequence of the magnetic layers on DMI.

Ultrathin films, typically only a few atomic layers thick, are conventionally termed two-dimensional (2D) or quasi-two-dimensional (quasi-2D) systems. In particular, in a film with a thickness of one or two atomic layers, the unit cell of both face-centered cubic (fcc) and body-centered cubic (bcc) lattices is not complete and the electronic and magnetic interaction are confined in the film plane. It is, therefore, of great fundamental  interest to understand the dimensionality effects on DMI in such quasi 2D systems.

It is worth mentioning that addressing the aforementioned  points requires an appropriate experimental tool capable of quantitatively probing DMI. While real-space imaging of spin textures can be used to indirectly quantify DMI (in particular the effective or the so-called micromagnetic DMI, see Ref.~\cite{Kuepferling2023} and references therein for details), magnon spectroscopy offers a direct quantitative method for probing this interaction. \cite{Zakeri2010,Zakeri2012}. The most straightforward way to precisely quantify the atomistic DMI is to probe both symmetric and antisymmetric interactions at atomic length-scales. This can be accomplished by probing high-wavevector magnetic excitations with wavelengths comparable to interatomic distances, as has been demonstrated in Refs. \cite{Tsurkan2020,Zakeri2023}. In this regard, a reasonably high energy, momentum, and spin resolution is essential  to probe tiny effects and provide quantitative values of $\vec{D}_{ij}$s.

Here we report on the impact of two critical factors on the atomistic DMI, i.e., (i) the number of $d$-state electrons of the interface atomic layer, and (ii) the number and the  sequence of the atomic layers.  With this, we unravel the impact of dimensionality on the atomistic DMI. We design specific magnetic architectures with desired specifications. We grow atomic bi- and trilayers of Fe, Co and Ni on the Ir(001) substrate and probe the atomistic DMI in such atomically designed structures. Utilizing Ir(001) as a substrate facilitates the growth of high-quality atomic multilayers. The near-invariant lattice structure and interatomic distances of the layered magnets grown upon this substrate are essential for accurately disentangling intrinsic effects from those arising from structural variations. We will discuss the role of the layer stacking of the ferromagnetic layers on the atomistic DMI.
It will be demonstrated experimentally, for the first time, that the number of the $3d$ states does not only influence the strength of DMI, but also its chirality. At first glance the general trend is that the intralayer atomistic  DMI in the interface atomic layer is directly correlated with the number of unpaired $3d$ states. However, due to the complex pattern of $\vec{D}_{ij}$s in each atomic layer the change in the effective DMI goes beyond a simple scaling law \cite{Kuepferling2023,Cho2015,Chaurasiya2016,LoConte2017,Belmeguenai2018,Ourdani2021}.  We will demonstrate that even large values of $\vec{D}_{ij}$ may result in a small effective DMI and vice versa. As another consequence of the complex pattern of the atomistic DMI, no simple prediction of the dependence of the effective DMI on the number of atomic layers is possible. Moreover, we will show that changing the number and the sequence of the atomic layers provide a versatile tool for tuning DMI. We anticipate that the results provide guidelines for designing complex topological spin textures in layered magnetic architectures, via tuning the atomistic DMI.

\begin{figure}[h!]
	\centering
	\includegraphics[width=1\columnwidth]{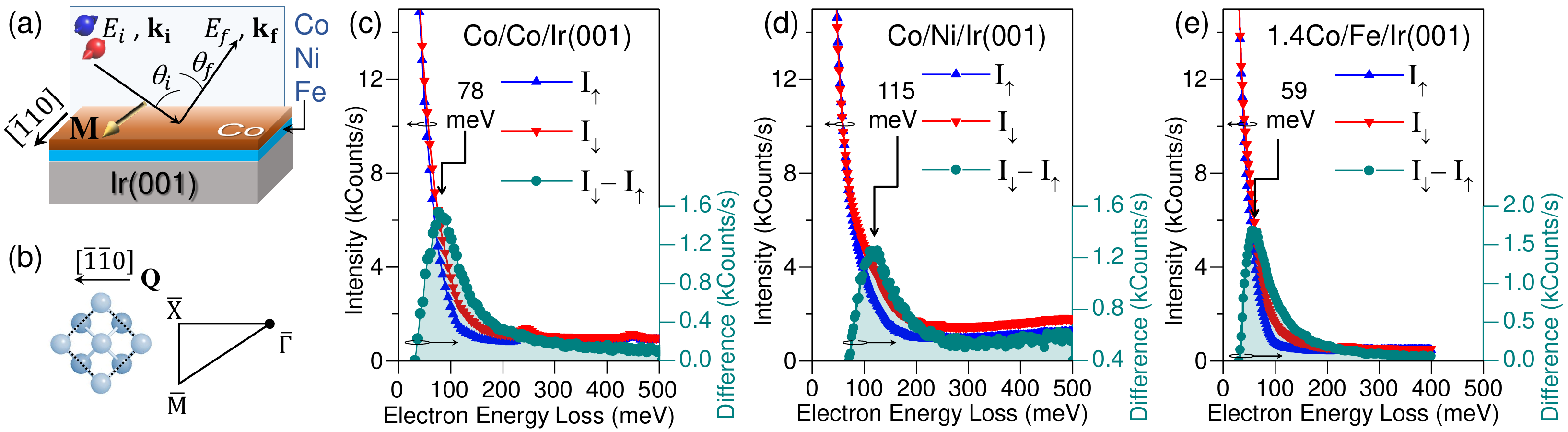}
	\caption{(a) The scattering geometry used for the SPHREELS experiments. $E_i$ ($E_f$),  $\vec{k_i}$ ($\vec{k_f}$) and $\theta_i$ ($\theta_f$) denote the energy, momentum and angle of the incident (scattered) electrons, respectively. (b) The magnon momentum $\vec{Q}$ with respect to the real and reciprocal space of the lattice. (c)--(e) Typical SPHREELS spectra recorded at a wavevector of $Q=0.6$~\AA$^{-1}$ on different multilayer systems epitaxially grown on Ir(001). The spectra were recorded at the incident energy of (c) $E_i=8$~eV, (d)  $E_i=10$~eV and (e) $E_i=8$~eV.  All spectra were recorded at room temperature. The spectra shown in red and blue, denoted by $I_{\downarrow}$  and $I_{\uparrow}$, were recorded with the spin polarization vector of the incident electron beam being parallel and antiparallel to the magnetization $\vec{M}$, respectively. The difference spectrum $I_{\downarrow}-I_{\uparrow}$ is shown by the sea-green color. In the ball model presented in (b), spheres of a lighter color denote atoms in the surface layer, while those of a darker color represent atoms in the layer below. For clarity, the conventional surface unit cell of the fcc(001) lattice (including two surface atoms)  is also shown by the dotted square. The magnon energy $\varepsilon (Q)$ is $78|^{+3}_{-1}$ meV, $115|^{+3}_{-1}$ meV, and $59|^{+2}_{-2}$ meV in (c), (d) and (e), respectively.}
	\label{Fig1:Spectra}
\end{figure}

\section{Results and Discussion}

\subsection{Quantification of the atomistic DMI}

Fe, Co, and Ni appear sequentially in Period 4 of the periodic table, with their $3d$ orbitals being progressively filled. This is also true in the solid form, where the $sp$ states hybridize with or overlap the $d$ states. In order to address the impact of the number of $3d$ electrons on the DMI we primarily focus on single atomic layers of Co, Ni and Fe epitaxially grown on Ir(001).  These monolayer magnets are covered with 1 or 1.4 monolayer of Co. In the following we refer these systems to as Co/Co/Ir(001), Co/Ni/Ir(001), and 1.4Co/Fe/Ir(001), respectively. The atomistic DMI was probed by spectroscopy of collective magnetic excitations (magnons, see Section~\ref{Sec:ExperimentalSection} for details of sample preparation and the  spectroscopy experiments). The experiments were performed on atomically designed ultrathin films and multilayres grown on atomically flat surface of an Ir(001) single crystal. The magnons were probed by means of spin-polarized high-resolution electron energy-loss spectroscopy (SPHREELS) \cite{Zakeri2013a, Zakeri2013, Zakeri2013b,Zakeri2014}.
The scattering geometry used in our experiments is shown in Figure~\ref{Fig1:Spectra}(a). The scattering plane was adjusted so that the magnons momentum $\vec{Q}$ was parallel to the $\bar{\Gamma}$--$\bar{\rm X}$ direction, as indicated in Figure~\ref{Fig1:Spectra}(b). Typical SPHREEL spectra recorded on such layered structures are presented in Figures~\ref{Fig1:Spectra}(c)--(e).  $I_{\downarrow}$ ($I_{\uparrow}$) represents the SPHREEL spectra when the spin polarization of the incoming electron beam is parallel (antiparallel) to the ground state magnetization $\vec{M}$. The difference spectrum $I_{\downarrow}-I_{\uparrow}$ includes all the necessary information regarding the magnons, e.g., their energy and lifetime \cite{Zhang2011, Zhang2012}.  In the data shown in Figures~\ref{Fig1:Spectra}(c)--(e) $\vec{M}$ was parallel to the $[\overline{1}10]$-direction and $\vec{Q}$ was along the $[\overline{1}\overline{1}0]$-direction. The difference spectra were evaluated using a Lorentzian function describing the intrinsic magnon signal, convoluted with a Gaussian function, accounting for the experimental resolution (see for example Refs. \cite{Zhang2011, Zhang2012, Zakeri2021a} for details). The magnon excitation energy $\varepsilon(Q)$ is given by the peak position of the Lorentzian. Our analysis reveals that $\varepsilon(Q)$ is $78|^{+3}_{-1}$ meV, $115|^{+3}_{-1}$ meV and $59|^{+2}_{-2}$ meV for the spectra shown in Figures~\ref{Fig1:Spectra}(c), (d) and (e), respectively.

\begin{figure}[h!]
	\centering
	\includegraphics[width=0.90\columnwidth]{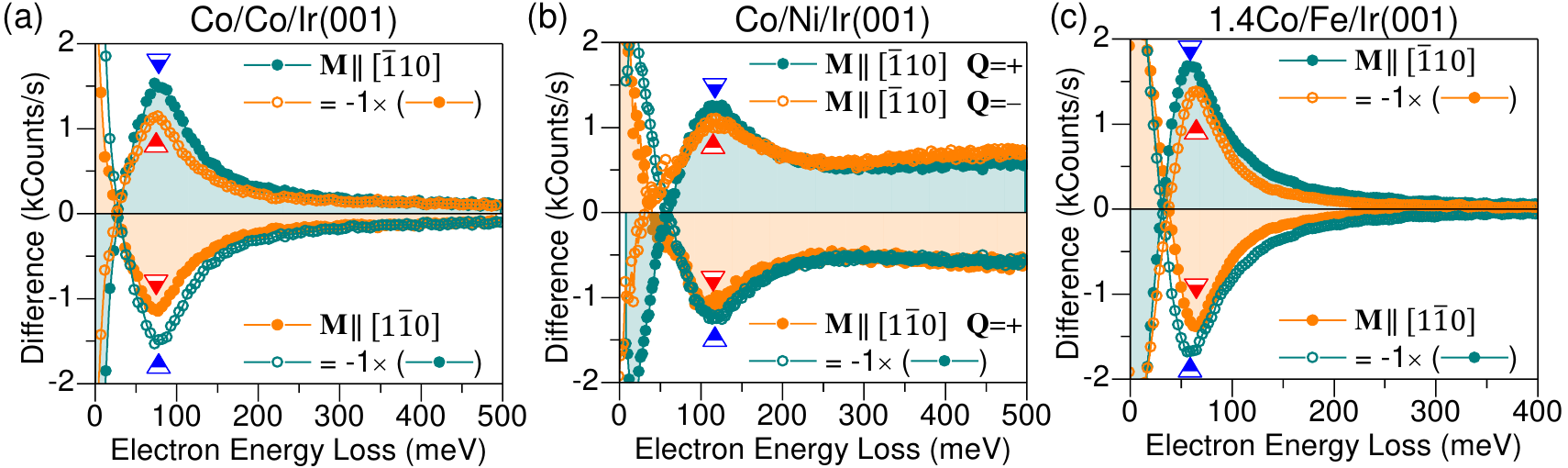}
	\caption{The difference spectra recorded on different layered systems. The spectra were recorded at $Q=+0.6$~\AA$^{-1}$ for the two possible directions of magnetization, i.e.,  $\vec{M}\parallel[\overline{1}10]$ (seagreen solid circles) and $\vec{M}\parallel[1\overline{1}0]$  (orange solid circles). In order to facilitate the comparison of the spectra with different magnetization directions, the same spectra multiplied by $-1$ are shown as well. In the case of Co/Ni/Ir(001) the data for both directions of $\vec{Q}$ and $\vec{M}$ are shown. For the cases denoted with $\vec{Q}=+$ and $\vec{Q}=-$ the propagation direction is along the [$\overline{1}\overline{1}$0] and [110] directions, respectively. The blue and red marks indicate the magnon energies $\varepsilon(Q)$. The corresponding values of $\Delta \varepsilon(Q)$ are $3|^{+3}_{-1}$ meV, $2|^{+3}_{-1}$ meV and $-6|^{+2}_{-4}$ meV, for Co/Co/Ir(001), Co/Ni/Ir(001)  and 1.4Co/Fe/Ir(001) systems, respectively.}
	\label{Fig2:Difference}
\end{figure}

The presence of  DMI leads to an asymmetry in the magnon dispersion relation. Hence, this interaction can be measured by probing the magnon energy asymmetry  $\Delta\varepsilon(Q)=\varepsilon(Q)-\varepsilon(-Q)$ as a function of the magnon momentum $Q$ \cite{Udvardi2009,Zakeri2010,Costa2010a,Zakeri2012,Zakeri2017,Tsurkan2020,Costa2020}. Probing $\Delta\varepsilon(Q)$ can be accomplished in two ways; either by changing the scattering geometry (which directly switches the direction of $\vec{Q}$) or by switching the sample magnetization $\vec{M}$ to the opposite direction. In Figure \ref{Fig2:Difference} we show the difference spectra recorded for $Q=\pm0.6$~\AA$^{-1}$ and for opposite orientations of $\vec{M}$. The high momentum resolution of the spectrometer manifests itself in the measurements performed for $\pm Q$, compared to those recorded by filliping the sample magnetization. The best examples are the spectra recorded on the Co/Ni/Ir(001) system presented in Figure~\ref{Fig2:Difference}(b), in which one clearly observes that the spectrum for $\vec{Q}=-$ and $\vec{M} \parallel[\overline{1}10]$ is almost identical to $\vec{Q}=+$ and $\vec{M} \parallel[1\overline{1}0]$. Even the absolute value of the intensity (count rates) is very similar.
Th data shown in Figure~\ref{Fig1:Spectra} unambiguously indicate the presence of an energy asymmetry for all three systems. $\Delta\varepsilon$ seems to be the largest for the case of 1.4Co/Fe/Ir(001) and smallest for the case of Co/Ni/Ir(001). More interestingly, the sign of $\Delta\varepsilon$ is reversed in the  case of 1.4Co/Fe/Ir(001) compared to the Co/Ni/Ir(001) and Co/Co/Ir(001) systems. The opposite chirality of DMI in 1.4Co/Fe/Ir(001) is manifested by the inverted sign of $\Delta\varepsilon$ compared to the systems having either a Co or a Ni interface layer.

\begin{figure}[h!]
	\centering
	\includegraphics[width=0.6\columnwidth]{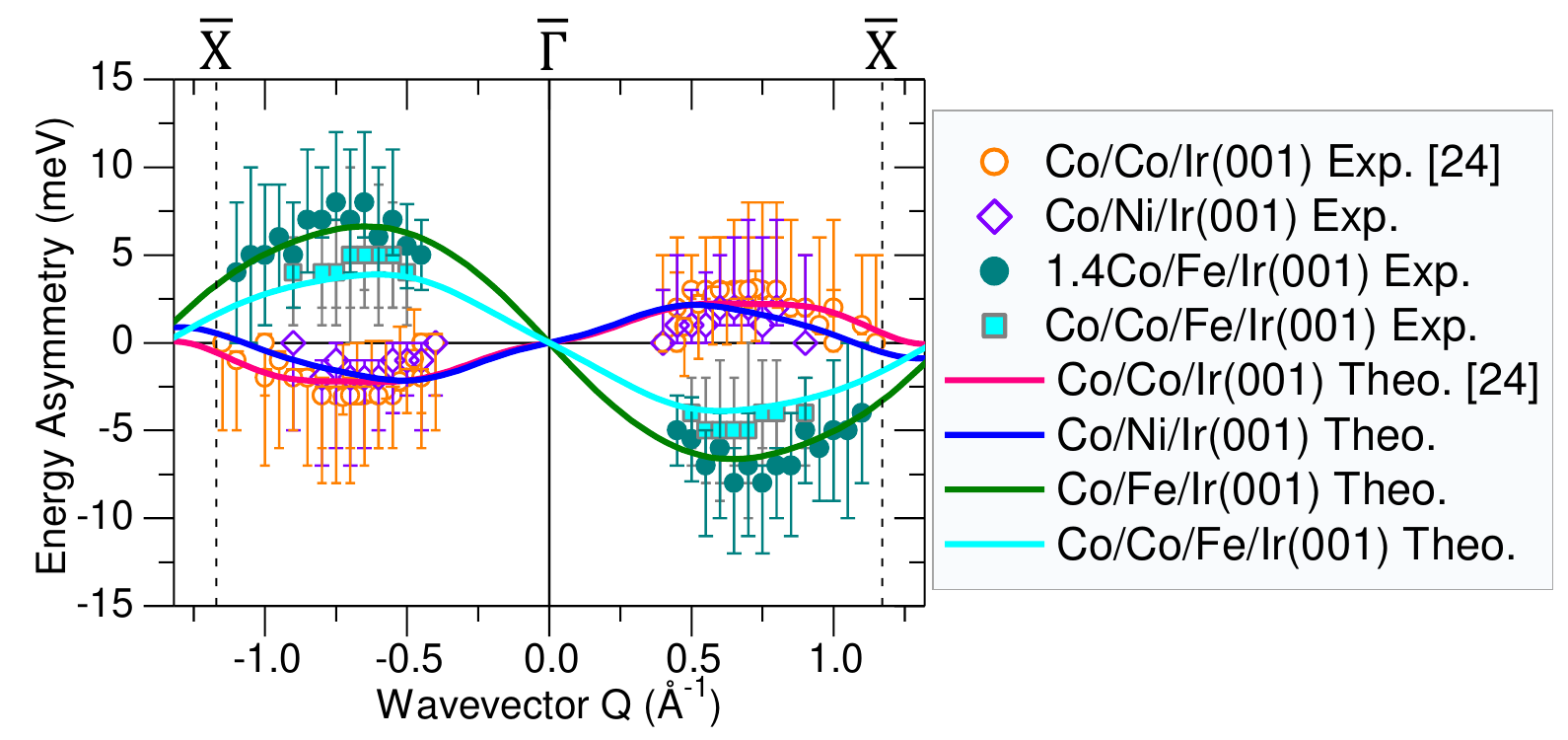}
	\caption{Dispersion relation of the DM energy $\Delta\varepsilon(Q)$ for different layered structures. The experimental data are shown by the symbols and the results based on the \textit{ab initio} calculations are shown by solid curves. Different colors indicate different systems. The opposite chirality of DMI in the 1.4Co/Fe/Ir(001) and Co/Co/Fe/Ir(001) samples is manifested by the inverted sign of $\Delta\varepsilon(Q)$ compared to the other systems. The data for the Co/Co/Ir(001) are reprinted with permission from \cite{Zakeri2023a}. Copyright (2023) by the	American Physical Society.}
	\label{Fig3:DMIDispersion}
\end{figure}

In order to gain more insights into the physics of the atomistic DMI in these systems we probed $\Delta\varepsilon$ as a function of $Q$ and the results are summarized in Figure~\ref{Fig3:DMIDispersion}. The experimental data of $\varepsilon (Q)$ (the magnon dispersion relation) for different systems are presented in Supporting Figure~S1 of the Supporting Information. By analogy, $\Delta \varepsilon(Q)$ can be regarded as the dispersion relation of DM energy, as it shows the energy landscape of DMI in momentum space. This quantity provides access to the atomistic DMI without the need for probing other magnetic parameters. Looking at the data presented in Figure~\ref{Fig3:DMIDispersion} one can conclude that the DMI at the Co/Ir interface is larger than that at the Ni/Ir interface. Both interfaces possess a smaller DMI compared to the Fe/Ir interface.  We will see that this is not the only important factor and the atomistic DMI depends critically on other factors, e.g., the number and sequence of the atomic layers.

In order to understand the experimental data in more details we performed first-principles calculations of the magnetic properties of the experimentally investigated layered structures. The calculations are based on the fully relativistic Korringa-Kohn-Rostoker electronic structure method (see Section~\ref{Sec:ExperimentalSection} for the technical details of the calculations).
The output was used to calculate  $\Delta\varepsilon(Q)$ \cite{Zakeri2010,Tsurkan2020,Zakeri2023a}. The results are presented in Figure~\ref{Fig3:DMIDispersion} and are compared to those of the experiments.

It is worth pointing out that in the first-principles calculations the variations in the interlayer distances have been fully taken into account. This is accomplished by using the experimental interatomic distances as the structural inputs of the calculations.
Hence, one source for the minor deviation of the experimental results from those of the calculations could be that the Ir(001) surface tends to reconstruct and form a 1$\times$5 reconstruction. The reconstruction appears usually in two mutually orthogonal domains. Although the experiments were performed on flat Ir(001) surfaces, it could be that in the initial stage of growth some small domain of  Ir(001)--1$\times$5 are formed.
Another source of deviation might be marginal interdiffusion of the atomic layers at interfaces. Although the samples are of high-quality, the real samples measured in the experiment are never perfect and ideal, as in the calculations.

The agreement between the numerical calculations and the experimental results confirms the reliability of the first-principles calculations. It is, therefore, useful to analyze $\vec{D}_{ij}$ in detail.

\begin{figure}[h!]
	\centering
	\includegraphics[width=0.5\columnwidth]{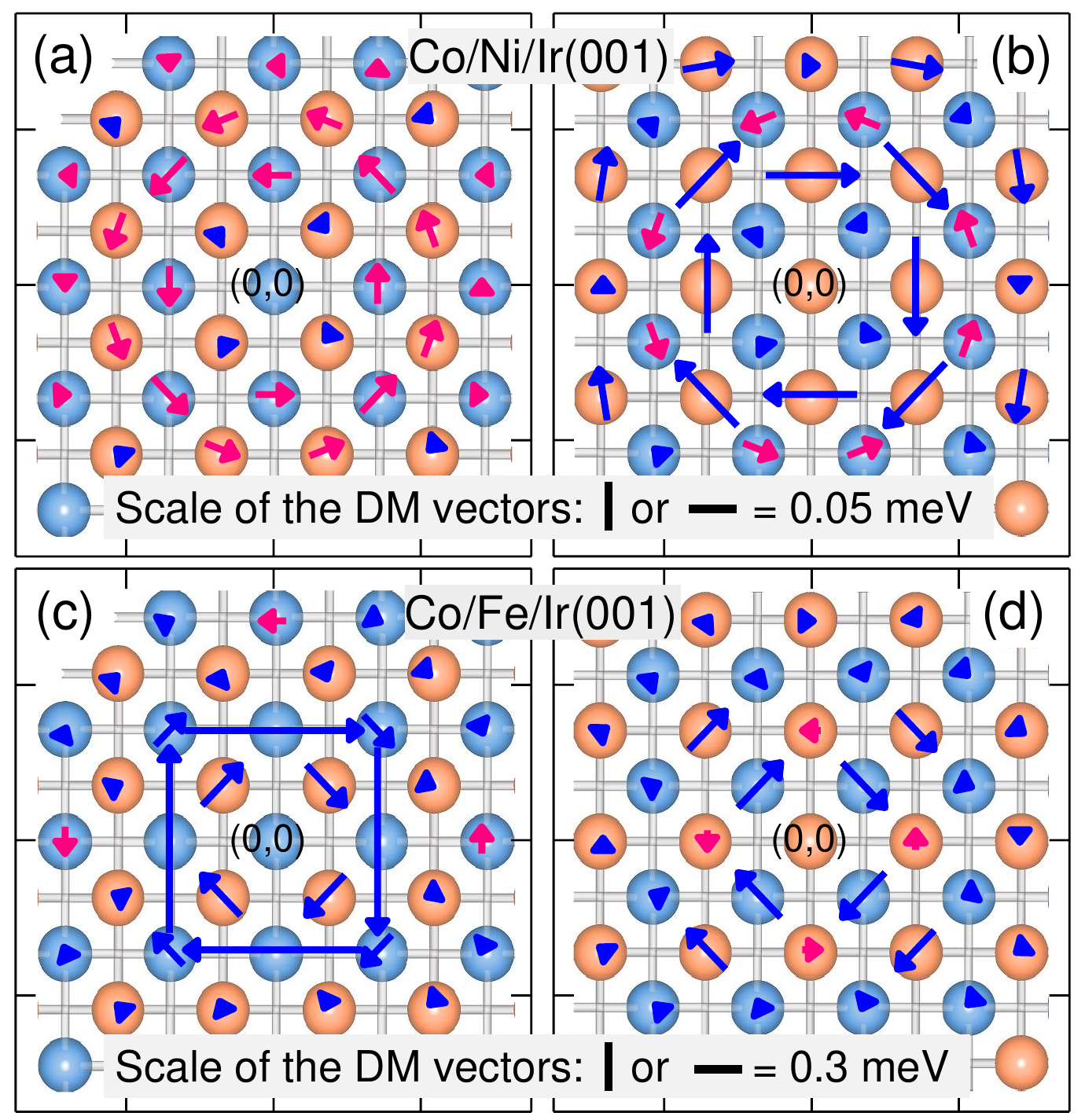}
	\caption{The calculated atomistic DM vectors for [(a) and (b)] Co/Ni/Ir(001) and [(c) and (d)] Co/Fe/Ir(001). The results for the case in which the origin site is located within the interface layer are shown in (a) and (c) and those for the case in which the origin site is located in the surface layer are shown in (b) and (d).  In the ball representation the surface atomic layer is shown by the orange color and the interface layer is shown in light-blue color. The color of $\vec{D}_{ij}$s  indicates their chirality (red for counter clockwise and blue for clockwise). The scales of $\vec{D}_{ij}$s are provided for each systems separately.}
	\label{Fig4:DMvectors}
\end{figure}

In Figure~\ref{Fig4:DMvectors} the layer-resolved $\vec{D}_{ij}$ is presented for the Co/Ni/Ir(001) and Co/Fe/Ir(001) systems. The results of the Co double layer on Ir(001) may be found in Ref.~\cite{Zakeri2023a}. The antisymmetric interaction of a Ni atom located at the origin $(0,0)$ within the interface layer, with the other Ni atoms located at a distance  $\vec{R}_{j}$ in the same layer (light-blue color)
or  the Co atoms located in the surface layer (orange color) are shown in Figure~\ref{Fig4:DMvectors}(a). The DM vectors describing the interaction between a Co atom located at $(0,0)$ in the surface layer with the surrounding atoms are presented in Figure~\ref{Fig4:DMvectors}(b). The chirality of DM vectors is denoted using different colors (red for a counter-clockwise rotation and blue for a clockwise rotation of the $\vec{D}_{ij}$s. The sign of the chirality index of the DMI is directly given by the direction of the $\vec{D}_{ij}$s.
The results of the Co/Fe/Ir(001) system are presented in Figures~\ref{Fig4:DMvectors}(c) and (d). In a similar way the DM vectors representing the interaction between an Fe atom, located at the origin, with all the neighboring atoms are shown in Figure~\ref{Fig4:DMvectors}(c) and those representing the interaction of the middle Co atom, located at (0,0) within the surface layer, with the other atoms are shown in Figure~\ref{Fig4:DMvectors}(d). One immediate conclusion of the data presented in Figure~\ref{Fig4:DMvectors} is that the DMI at the Ni/Ir interface is much smaller than at the Fe/Ir interface. For instance, the second nearest neighbor interaction which describes the interaction between neighboring Ni atoms within the interface Ni layer ($D_{2,x}=0.049$~meV, $D_{2,y}=0$~meV) is by a factor of 27 smaller than the corresponding interaction within the Fe interface atomic layer ($D_{2,x}=-1.33$~meV, $D_{2,y}=0$~meV). Likewise, the interlayer DMI of Ni--Co ($D_{1,x}=-0.0018$~meV, $D_{1,y}=+0.0018$~meV) is much weaker compared to that of  Fe--Co ($D_{1,x}=-0.29$~meV, $D_{1,y}=+0.29$~meV), even though the interatomic distances of these two layers are the same. Looking at the DM vectors within the Co atomic layers one realizes that the nearest neighbor intralayer DMI in the Co layer on top of the Ni ($D_{2,x}=-0.124$~meV, $D_{2,y}=0$~meV) has nearly the same magnitude ($D_{2,x}=+0.126$~meV, $D_{2,y}=0$~meV). This observation that the atomistic DMI of the Fe/Ir interface is much larger than that of the Co/Ir interface has also been observed in the theoretical calculations by Yang \textit{et al.}, on a multilayer  stack composed of 3Ir$|$3Co$|$3Pt, all having a (111) surface orientation \cite{Yang2018}. It is important to notice that in our case all the layers possess a (001) surface orientation. It seems that the Fe/Ir interface exhibits a large atomistic DMI, irrespective of the surface orientation. Another very important result is that the chirality of the $\vec{D}_{ij}$s in the Co/Ni/Ir(001) and Co/Fe/Ir(001) systems is very different. While within the Ni layer one can clearly see that the chirality is of counter-clockwise type, within the Fe atomic layer it is of clockwise nature. In addition, the pattern of  $\vec{D}_{ij}$s within the Co layers for the two systems is different. Inside the surface Co layers the pattern is a mixture of both chiralities. Comparing the results to those of the Co double layer grown on the same surface indicates that the change in the pattern of $\vec{D}_{ij}$s in the surface Co layer is a consequence of the reconstruction of the electronic structure of the system as a result of replacing the interface layer with a different element.
The finding that $\vec{D}_{ij}$s within the interface Fe layer possess an opposite chirality compared to those within the interface Ni atomic layer is experimentally supported by the opposite sign of the energy asymmetry when comparing the experimental results of Co/Fe/Ir(001) to those of Co/Ni/Ir(001), presented in Figure~\ref{Fig3:DMIDispersion}. The effect is far beyond the error bars. As one clearly observes, for the samples with an Fe atomic layer at the  interface $\Delta\varepsilon(Q)$ is negative for positive $Q$ and positive for negative $Q$. This behavior stands in opposition to that observed in samples, where Co or Ni is present at the interface. Likewise, the magnitude of DMI is somehow encoded in the absolute values of $\Delta\varepsilon(Q)$. These values are larger for the case of Co/Fe/Ir(001) compared to the other systems. It is important to emphasize that not only the absolute values of  $\Delta\varepsilon(Q)$  but also its behavior as a function of $Q$, are crucial for the determination of DMI, given the complex pattern of atomistic DMI in layered structures.

Additionally, epitaxial relationships and lattice mismatch significantly influence magnetic parameters, particularly atomistic DMI. The structural similarity and comparable interatomic distances of ultrathin Fe, Co, and Ni films on Ir(001) offer a key experimental advantage. By ensuring that the replacement of the interface layer does not influence the epitaxial strain, one can isolate and observe the effects associated with variations in the unoccupied $3d$ states. we would like to emphasize that in the first-principles calculations  the experimental lattice parameters were used. Given the fact that all multilayer structures within this study exhibit identical geometrical structures, the observed variations in the atomistic DMI can be confidently attributed to changes in the number of electronic  $d$ states.

\begin{figure*}[h!]
	\centering
	\includegraphics[width=0.7\columnwidth]{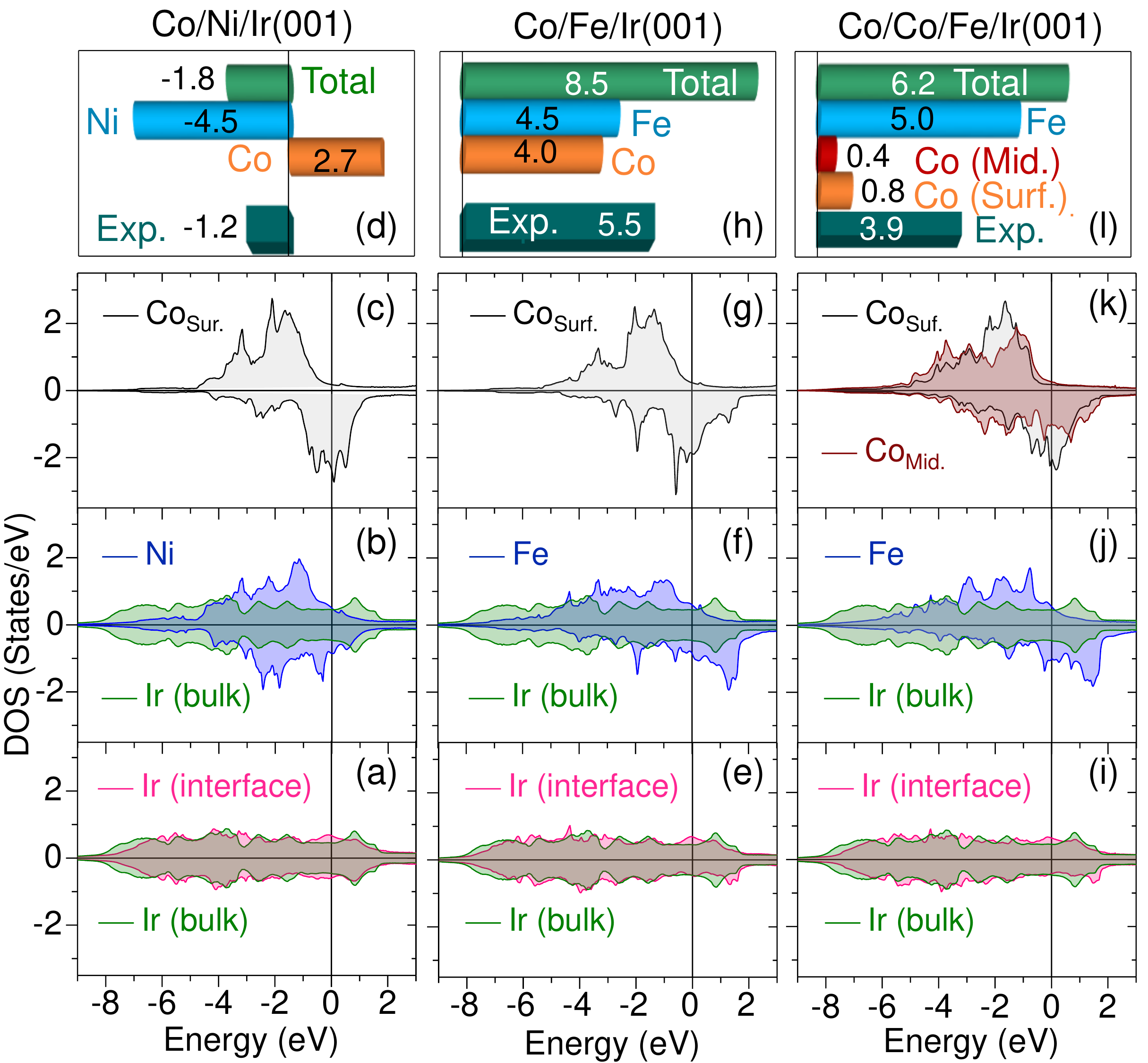}
	\caption{Spin and layer-resolved DOS and $\mathcal{D}$ of different layered structures. The results of Co/Ni/Ir(001), Co/Fe/Ir(001) and Co/Co/Fe/Ir(001) are presented in (a)--(d), (e)--(h) and (i)--(l), respectively. The values of $\mathcal{D}$ are given in meV\AA/$\mu_B$. The experimental results shown in (h) are for 1.4Co/Fe/Ir(001). In all the DOS representations the Fermi-level is set to zero and is indicated by the solid vertical lines.}
	\label{Fig5:DOS}
\end{figure*}

\subsection{Interplay between orbital occupancy and the atomistic DMI}

Since DMI requires a broken inversion symmetry in combination with a large spin-orbit coupling, it is therefore straightforward to investigate the spin-orbit energy in the system in detail, as has been done in Ref.~\cite{Yang2015}. In the case of interfacial DMI the spin-orbit coupling is supported by the nonmagnetic substrate (in our case Ir(001)). Because all films were grown on the Ir(001) substrate with nearly identical geometrical parameters, including surface orientation and interatomic distances, the spin-orbit coupling strength acting on the layers is assumed to be identical for all systems. The number of electronic $d$-states were varied by changing the interface atomic layer, keeping the surface layer unchanged.
Therefore, in our analysis we focus on the changes in the electronic states rather than changes in the spin-orbit energy.
To this end, it is very useful to perform an analysis of the features of the density of states (DOS) near the Fermi level that are important for the magnetic interactions \cite{Chuang2012}. As pointed out, an important consideration is that due to the antisymmetric nature of DMI it involves the spin-orbit spin-mixed states, mainly supported by the substrate. Hence, places inside the electronic DOS, which can lead to the appearance of such states are of particular importance.

In Figure~\ref{Fig5:DOS} the spin and layer-resolved DOS of different layered structures are presented. The analysis reveals important features in the DOS. We first focus on the results of the Co/Ni/Ir(001) system. A comparison between the DOS of the interface Ir atoms compared to that of the Ir atoms sitting deeply inside the substrate is provided in Figure~\ref{Fig5:DOS}(a), indicating that only the majority spin states are slightly modified. In particular the unoccupied states slightly below 1~eV are shifted towards the Fermi level. A detailed analysis of these states indicates that they are mainly of $d_{xz}$, $d_{yz}$ and $d_{x^2-y^2}$ character (see Supporting Note~S1 and Supporting Figure~S2 of Supporting Information). These $5d$ states of Ir are responsible for the electronic  hybridization. The energetic overlap of these states with the electronic states of Ni, shown in Figure~\ref{Fig5:DOS}(b), is rather small. This is due to the fact that in Ni a large fraction of the electronic states are occupied and hence are located below the Fermi level \cite{Zakeri2024a}. The smaller overlap of the $3d$ states of the Ni with the $5d$ states of Ir leads, obviously,  to a small DMI in Ni. This interpretation is in agreement with the Hund's first rule as the main mechanism responsible for the hybridization and the DMI \cite{Belabbes2016,Zhu2022}.  The partial DOS of the Co atoms shown in Figure~\ref{Fig5:DOS}(c) suggests that, in this case, a moderately high degree of hybridization is expected due to the presence of a sufficient number of  Ir $5d$ states degenerated with the unoccupied minority states of Co.

In order to be able to compare the values of  $\vec{D}_{ij}$s of different layers and to that of the experimental magnon energy asymmetry we define the quantity $\mathcal{D}$ given by $\mathcal{D}=\sum_{ij} (1/\mu_i)\vec{D}_{ij}\cdot \vec{R}_{ij}$. The layer-resolved $\mathcal{D}$ is shown in the upper rows of Figure~\ref{Fig5:DOS} for each system. The results of Co/Ni/Ir(001) indicate that $\mathcal{D}$ within the Ni (Co) layer is negative (positive) as a result of the counter-clockwise (clockwise) rotation of $\vec{D}_{ij}$ [see Figure~\ref{Fig5:DOS}(d)]. Due to the larger contribution of the Ni layer, the value of total  $\mathcal{D}$ is negative in agreement with the experimental value.

The DOS of Co/Fe/Ir(001) is presented in Figures~\ref{Fig5:DOS}(e)--(g). As apparent from Figure~\ref{Fig5:DOS}(e) the shift in the Ir $5d$ states is significant. In particular the unoccupied states of the interface Ir atoms are shifted to higher energies (with respect to the bulk Ir) as a result of a high degree of hybridization. The most important observation is the presence of a considerably large amount of unoccupied states above the Fermi level in the Fe layer [see Figure~\ref{Fig5:DOS}(f)], which energetically degenerate with the spin-orbit mixed Ir $5d$ states. These states are predominantly of $d_{z^2}$, $d_{yz}$, $d_{xz}$ and $d_{xy}$ and partially of $d_{x^2-y^2}$, character \cite{Zakeri2013a,Chuang2014} (see  Supporting Figure~S2 of Supporting Information). Likewise, the Co surface layer exhibits a considerable amount of unoccupied $d$ states above the Fermi level  [see Figure~\ref{Fig5:DOS}(g)]. These states are extended to higher energies compared to the case of Co surface layer of the Co/Ni/Ir(001) system shown in Figure~\ref{Fig5:DOS}(c). This is a consequence of  the hybridization of these states with (i) the states of the Fe layer and (ii) those of the Ir interface layer. Both of these facts lead to a notable DMI in the Co layer. As a result, this system exhibits a rather large total $\mathcal{D}$ as observed in Figure~\ref{Fig5:DOS}(h).

Since the Fermi level is mainly determined by the substrate, when an additional Co layer is added on top the features above the Fermi level remains almost unchanged  [see Figures~\ref{Fig5:DOS}(i)--(k)]. However, due to the fact that different symmetries of orbitals are involved in the coupling \cite{Yang2015,Kim2018, Zakeri2024,Faber2024} the pattern of $\vec{D}_{ij}$s changes dramatically leading to a reduction of $\mathcal{D}$ in the Co layer, while the intralayer $\vec{D}_{ij}$s within the Fe and middle Co layer remain unchanged (see the discussion under Section~\ref{Sec:Dimesnionality} and Figure~\ref{Fig6:CoCoFe}). The reduction of $\mathcal{D}$ in Figure~\ref{Fig5:DOS}(l) is also a result of the strong coupling of the two Co layers which leads to a shift of the spin-density towards the surface layer, in line with the large magnetic moment of the middle Co  layer.  A detailed discussion of the spin-, orbital- and layer-resolved DOS is provided in Supporting Note~S1 and Supporting Figure~S2 of Supporting Information.

\subsection{The impact of dimensionality on  the atomistic DMI}\label{Sec:Dimesnionality}

In order to address the impact of dimensionality on $\vec{D_{ij}}$s we investigated the pattern of $\vec{D_{ij}}$s when increasing the number of atomic layers.  Since one monolayer of $3d$ ferromagnets on Ir(001) (and many other substrates) does not exhibit a ferromagnetic order, we investigated the effect in the transition between the second and third atomic layers. This is where the unit cell of fcc lattice becomes complete. As a side remark, the absence of a ferromagnetic order does not mean that the atomistic DMI in an atomic layer is inactive. In fact the atomistic DMI in such monolayers
is rather strong \cite{Bode2007,Heinze2011,Herve2018}. The ferromagnetic order and DMI are different physical aspects. 	In the SPHREELS experiments it is much easier to investigate samples which show a ferromagnetic order. In such
	a case one can take advantage of ``spin resolution'' by measuring the difference spectra, as shown in Figure~\ref{Fig2:Difference}. With this one can
	eliminate the effects associated with spin-independent processes, e.g., phonons or non-spin-flip processes, during the
	scattering event \cite{Qin2017}.

\begin{figure*}[t!]
	\centering
	\includegraphics[width=0.9\columnwidth]{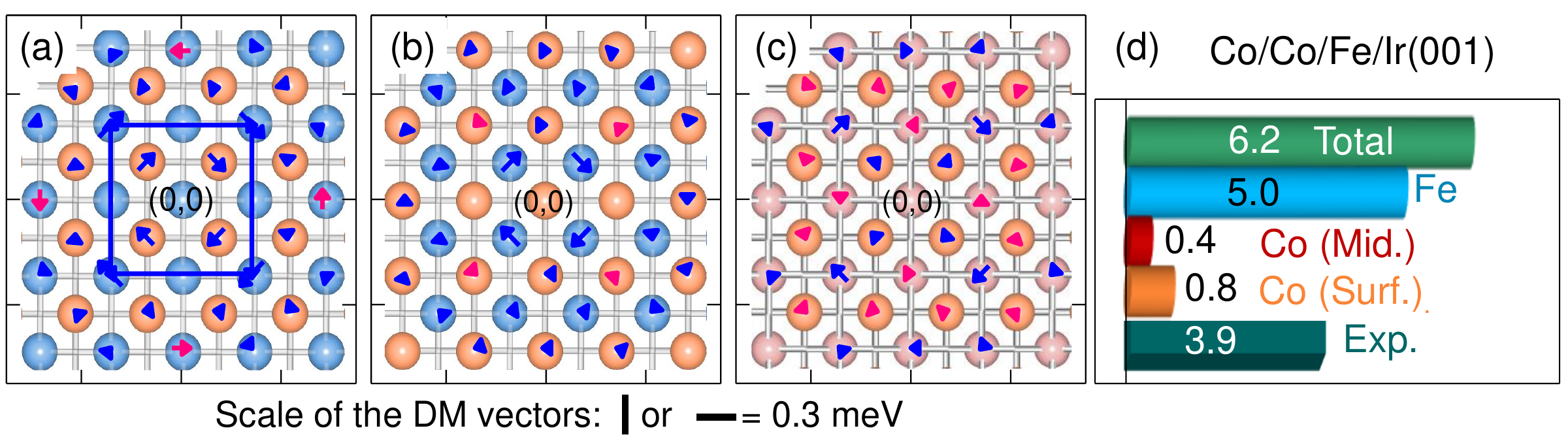}
	\caption{The calculated layer-resolved atomistic DM vectors  $\vec{D}_{ij}$  and the effective DMI $\mathcal{D}$ for the Co/Co/Fe/Ir(001) layered structure. (a)--(c) The atomistic DM vectors within each atomic layer of the Co/Co/Fe/Ir(001) layered structure. The results for the case in which the origin site is located within the interface Fe layer, the middle Co layer and the surface Co layer are shown in (a), (b) and (c), respectively. For the sake of simplicity only the interaction between the neighboring layers are shown.  The color of $\vec{D}_{ij}$s  indicates their chirality (red for counter-clockwise and blue for clockwise).  (d) The layer-resolved $\mathcal{D}$ given in meV\AA/$\mu_B$.}
	\label{Fig6:CoCoFe}
\end{figure*}

The results of the $\vec{D_{ij}}$s of the Co/Co/Fe/Ir(001) structure are presented in
Figure~\ref{Fig6:CoCoFe}. Comparing the results to those of Co/Fe/Ir(001), shown in Figures~\ref{Fig4:DMvectors}(c) and \ref{Fig4:DMvectors}(d), indicates that the values of $\vec{D_{ij}}$s inside the Co layers are very different. The dramatic changes of $\vec{D}_{ij}$s and, consequently, the reduction of $\mathcal{D}$ in the Co layers, while the intralayer $\vec{D}_{ij}$s within the Fe layer have remained unchanged, is a consequence of a change in the dimensionality of the system. When the second Co layer is added it is strongly coupled to the Co layer below. This fact leads to a shift of the spin-density towards the surface layer, in line with the large magnetic moment observed for the surface Co layer. Before adding the Co layer the spin-density is mainly confined in the Co and Fe layers. The presence of the additional Co layer leads to a redistribution of the electrons toward the third direction, i.e., the direction perpendicular to the surface. Such an effect has been observed to substantially change the overlap of the electronic states and, consequently, change the symmetric exchange interaction, leading to a nonmonotonic behavior of the Heisenberg exchange interaction as a function of film thickness \cite{Zhang2010, Chuang2012}. A similar but much more complex  effect is also expected here. The complexity is, partially, due to the fact that the spin-orbit coupling is mainly supported by the presence of the substrate and, hence, does not affect all layers equally. We note that all these have  properly been taken into account in our first-principles calculations.

Our results predict that $\varepsilon(Q)$ shall be strongly suppressed when exchanging the interface Fe layer and the middle Co layer (see Supporting Figure~S3 of Supporting Information). However, the strength of $\mathcal{D}$ is found to be governed by $\vec{D}_{ij}$s within the Co layers. This is due to the fact that the orbitals essential for the hybridization are supported by the Fe middle layer. Our experiments indicate that  $\varepsilon(Q=0.6)\leq1$~meV, confirming the validity of our discussions.

As another prove of the impact of the dimensionality on the atomistic DMI we have performed calculations for one and two atomic layers of Co grown on the Ir(001) surface and observed the same effect. The main results of this investigation are as follows.

In the case of Co/Ir(001) it was observed that $\vec{D}_{ij}$s are large. For instance we obtained the values of ($D_{1\parallel},x$ = $-1.6$, $D_{1\parallel},y = 0$) and ($D_{2\parallel},x = D_{2\parallel},y = +0.13$)~meV, representing the first and second nearest neighbors sitting in the same atomic plane, respectively. In the case of Co/Co/Ir(001) these values were found to be ($D_{1\parallel},x = -0.45$, $D_{1\parallel},y = 0$) and ($D_{2\parallel},x = D_{2\parallel},y = +0.1$)~meV for the interface Co layer  and ($D_{1\parallel},x = -0.14$, $D_{1\parallel},y = 0$) and ($D_{2\parallel},x = D_{2\parallel},y = +0.16$)~meV for the surface Co layer. This is again due to the fact that adding the second Co layer changes the dimensionality of the system. This effect in the electronic structure appears mainly as a  shift of the spin-density in the direction perpendicular to the surface, as observed by analysis of the spin-resolved DOS.

The effects mentioned above are direct consequences of the dimensionality of the systems. The concept of dimensionality is also important in the context of newly discovered materials, known as 2D van der Waals (vdW) magnets. Since the electronic and magnetic interactions within each vdW layer in this class of materials are much stronger than those between the vdW layers, one would expect to observe similar effects there.

\section{Conclusion}

In conclusion, we examined the impact of two critical factors on the atomistic chiral magnetic interaction, i.e., (i) the density of the electronic $d$-states of the interface atomic layer, and (ii) the number and the  sequence of the magnetic layers. It was demonstrated both experimentally and theoretically that the number of the unoccupied $3d$ states of the interface atomic layer is decisive for  both the magnitude and the chirality of the DM vectors. The general trend is that the magnitude of the intralayer atomistic  DMI in the interface layer is directly correlated with the number of unpaired $3d$ states within this layer. However, the pattern of $\vec{D}_{ij}$s inside this layers (and the layers on top) is very complex. Hence, a prediction of the total DMI goes beyond a simple scaling law. This fact was explained based on the complexity of the electronic structure and the contributions of different orbitals to the hybridization and the antisymmetric exchange interaction. This means that changing the number and the sequence of the atomic layers provides a way of tuning DMI in layered structures.
For instance, as it is demonstrated  for the layered structures with an atomic layer of Ni at the interface one obtains a rather weak DMI. If the idea is to slightly enhance DMI one can replace this layer with a Co layer. Likewise, if the idea is to change the chirality index (sign) of DMI one may use an Fe layer at the interface. In a similar way one can take advantage of the number and sequence of the layers in order to tune DMI. It is important to notice that there is no simple strategy to control the magnitude and chirality of DMI.
However, if the idea is to design a certain type of topological spin texture, one may use an inverse design strategy. For that one needs to estimate the optimized material's parameter. Once these parameters are estimated one can design the system by using the above mentioned knowledge and tuning the atomistic DMI. Our work demonstrates to which extent such a tuning of the atomistic DMI is possible.

As the final remark, while our primary focus was to address the atomistic DMI in Fe, Co, and Ni monolayers coupled to a Co layer grown on top, in principle, the study of monolayer systems does not necessarily require an additional atomic layer \cite{Prokop2009,Zakeri2017,Zakeri2014a,Qin2015}.  It would certainly be very interesting to probe DMI in such samples. However, single atomic layers of ferromagnetic elements grown on the Ir(001) surface (and
many other surfaces) do not exhibit a long-range ferromagnetic order at temperatures accessible using standard
He-flow cryostats. In such cases one needs to either investigate the spin-integrated signal or perform spin analysis after the scattering event (a spin-polarized beam and a spin-resolved detection). Such investigations are beyond the scope of the present study.

\section{Experimental Section}\label{Sec:ExperimentalSection}
\threesubsection{Sample preparation}

All the experiments were performed under ultrahigh vacuum conditions and at room temperature. First, the surface of the Ir(001) substrate was cleaned using a procedure described in details in Refs.~\cite{Zakeri2010a,Chuang2014,Chen2017,Zakeri2021,Zakeri2021a,Zakeri2023a}. Then, atomic layers of Fe, Co and Ni were grown by molecular beam epitaxy. Low-energy electron diffraction (LEED) recorded on the grown layered ferromagnets showed mainly a ($1\times1$) pattern and a well-ordered face-centered tetragonal structure of the multilayer structures \cite{Heinz2009,Tian2009,Kueppers1979}. In the first step, three different synthetic layered structures (mainly bilayers) were prepared and examined.

In order to unravel the impact of the number of $3d$ electrons on the DM vectors we first grow one atomic layer of Fe, Co and Ni on the Ir(001) surface. Since such monolayers do not exhibit a long-range ferromagnetic order one cannot take advantage of spin-resolution to probe the magnetic excitation (magnons) in these systems. One way to overcome this problem is to couple these atomic layers to another atomic layer on top. In this way the whole system will show a ferromagnetic order. Probing the acoustic magnon mode of the system provides an access to the magnetic interactions in the interface magnetic monolayer \cite{Zakeri2013a}. We, therefore, cover the Fe, Co and Ni monolayers with one atomic layer of Co. The choice of Co as the surface layer is based on the fact that it exhibits the highest excitation cross-section of magnons. The magnetic state of the samples was probed by means of the magneto-optical Kerr effect (MOKE) in the longitudinal geometry. The magnetization of the samples was found to lie in the plane for all layered structures. While the Co/Co and Co/Ni bilayers exhibit rectangular hysteresis loops at room temperature, the Co/Fe shows a S-shape loop. In order to obtain a well-ordered magnetic state for this case and avoid the effects associated with a possibly low Curie-temperature of this system we added an additional 0.4 monolayer of Co on this structure. The investigated system is referred to as 1.4Co/Fe/Ir(001).

In order to investigate the impact of the number and the sequence of the atomic layers on the DMI, in the next step, we also prepared Co/Co/Fe and Co/Fe/Co structures and compared the results to that of the magnetic bilayers.

The structural, chemical and morphological properties of $3d$ ferromagnetic elements on both the reconstructed and unreconstructed Ir(001) surface have been well investigated by means of IV analysis of low-energy electron diffraction (LEED) and atomic resolved scanning tunneling microscopy (STM) (see Refs. \cite{Faber2024, Heinz2009,Tian2009,Kueppers1979, Zakeri2024c}).
We performed a careful chemical and morphological investigation of our samples. The LEED patterns recorded on the bare Ir(001) and after each atomic layer deposition shows rather sharp LEED spots with no satellites, confirming an epitaxial growth of the films with a rather low surface roughness. Examples of LEED patterns recorded on some of our samples are presented in Supporting Figures~S4(a) and S4(b) of Supporting Information.

Moreover, we measured the intensity profile of the elastic diffuse scattering (EDS). Generally the EDS profile is strongly influenced by the presence of defects and the surface roughness. Since  in our experiments an incident beam energy of only 10~eV was used, the intensity profile of EDS shall reflect
the surface quality.
As an example we provide the EDS profile of the Co/Ni/Ir(001) sample in Supporting Figure~S4(c). The strong $Q$-dependent of the EDS profile is a clear signature of a high surface quality, a low surface roughness and relatively wide terraces. At $Q=0$ ($\bar{\Gamma}$--point) the intensity is very high, so that it cannot be measured in pulse-counting mode. The intensity drops by an order of magnitude while changing $Q$ from 0.05 to 0.15 \AA$^{-1}$, indicating a high sample quality.

\threesubsection{Probing magnon energy asymmetry}

The magnons were probed by means of spin-polarized high-resolution electron energy-loss spectroscopy (SPHREELS) \cite{Zakeri2013, Zakeri2014,Zakeri2018}. The SPHREEL spectra were recorded along the main symmetry direction of $\bar{\Gamma}$--$\bar{\rm{X}}$ of the surface Brillouin zone (SBZ). All the spectra were recorded at the magnetic remanent state. The incident electron energy was between 6 and 10 eV. The incident energy was chosen such that  the largest magnon excitation cross-section was observed. The energy resolution was in the range of 12--20 meV. The momentum resolution in our experiments was as high as $0.02$ \AA$^{-1}$. In the SPHREES spectra electrons with their spin parallel to the sample magnetization are referred to as minority electrons and those with spin polarization antiparallel to the sample magnetization are referred to as majority electrons.  The experiments were performed using a highly spin-polarized beam with a polarization of about 80\%. In the SPHREELS experiments magnons are excited by incidence of minority electrons, due to the conservation law of the total angular momentum. This fact leads to the appearance of a peak in the minority spin spectra ($I_{\downarrow}$) and, consequently, a peak in the difference spectra, defined as  $I_{\downarrow}-I_{\uparrow}$ \cite{Vollmer2003}.
Hence, the difference spectra include all the necessary information on magnons, e.g., their excitation energy $\varepsilon(Q)$ and lifetime. The magnon energies are extracted by fitting the difference spectra using a convolution of a Lorentzian and a Gaussian. The Lorentzian represents the actual magnon signal and the Gaussian shall represent the experimental resolution, which can be separately measured (see for example Refs. \cite{Zakeri2012,Zhang2012,Zakeri2021a,Qin2015, Zakeri2014,Qin2019}).

The quantity $\Delta \varepsilon(Q)$ is defined as $\varepsilon(Q)- \varepsilon(-Q)$. The most straightforward way to measure this quantity is to record the spectra for $Q$ and $-Q$ and, thereby, obtain $ \varepsilon(Q)$ and $ \varepsilon(-Q)$. The magnon wavevector is given by  $Q=|\vec{Q}|=Q_{\parallel} = k_{i} \sin\theta - k_{f} \sin(\theta_{0}-\theta)$,
where $k_{i}$ ($k_{f}$) is the magnitude of the wavevector of the incident (scattered) electrons, and  $\theta$ ($\theta_{0}$) is the angle between the incident beam and sample normal (the scattered beam). In order to probe magnons with positive and negative $Q$ one may change the incident and scattered angles. This may be achieved by rotating the sample about its main axis, i.e., changing $\theta$ and keeping $\theta_{0}$ fixed. Hence, any uncertainty in the determination of angles can lead to an uncertainty in $Q$ and consequently in $\varepsilon(Q)$. In order to fully eliminate this, one way would be to record the spectra at a fixed scattering geometry. Inverting the sign of $Q$ can be accomplished by reversing the sample magnetization (an extended discussion may be found in Ref.~\cite{Zakeri2010,Zakeri2011}). Due to the fact that reversing the direction of the magnetization is equivalent to a time-inversion experiment, it is also equivalent to inverting the sign of $Q$  (see also the data presented in Fig.~\ref{Fig2:Difference}). In order to ensure that this is practically true, we have performed both type of measurements. As it can be concluded  from the data presented in Fig.~\ref{Fig2:Difference} the possible errors cased by the scattering geometry on $\Delta \varepsilon(Q)$ are fairly small and negligible. This is due to the fact that the error in the scattering angles is very small (leading to a high momentum resolution). For the data presented in Fig.~\ref{Fig3:DMIDispersion} we use the measurement protocol in which  the opposite sign of $Q$ is achieved by reversing the sample magnetization. This means that in this set of data, the errors associated with the scattering geometry are entirely eliminated.

\threesubsection{\emph{Ab initio} density functional theory calculations of $\mathbf{D}_{ij}$s}

\emph{Ab initio} density-functional theory based calculations were performed for all the experimentally investigated layered structures.  Our first-principles calculations are based on the fully relativistic Korringa-Kohn-Rostoker electronic structure method \cite{Ebert2011}. We started with the self-consistent calculations of the electronic structure within the framework of a local density approximation of the density functional theory with the parametrization due to Vosko, Wilk and Nusair \cite{Vosko1980}. All the technical details of the scheme used to compute the symmetric and antisymmetric magnetic exchange interaction within the  framework of the magnetic force theorem \cite{Liechtenstein1987} are given in Ref.~\cite{Mankovsky2017}. A slab composed of eight atomic layers of Ir, ordered in fcc(001), followed by the $3d$ magnetic monolayers with another eight layers of vacuum on top of the magnetic layers, was embedded between the semi-infinite Ir substrate and semi-infinite vacuum region. The cut-off $l_{max} = 3$ was used for angular momentum expansion of the Green function, and the integration over 2D Brillouin zone was done using a $k$-mesh with $80\times80$ grid points in the 2D-BZ. In order to fully account for the geometrical structure the experimental interatomic distances were used as the input for the  first-principles calculations  \cite{Heinz2009,Chen2017,Zakeri2021a,Qin2019,Zakeri2021}. The in-plane interatomic distances were assumed to be the same as that of the Ir substrate. We used the values of Ir--Fe=Ir--Co=Ir--Ni=1.75~\AA~ and	Co--Co=Co--Fe=Ni--Co =1.6~\AA~ for the interlayer distances. The \textit{ab initio} calculations provide all the necessary magnetic parameters of the investigated systems, in particular the atomistic DM vectors $\vec{D}_{ij}$. Those values are shown in Figures~\ref{Fig4:DMvectors} and \ref{Fig6:CoCoFe}. The values of spin and orbital angular momentum were $\mu_S^{Co}=1.85$, $\mu_L^{Co}=0.135$, $\mu_S^{Ni}=0.34$ and $\mu_L^{Ni}=0.032$, in the case of Co/Ni bilayer, $\mu_S^{Co}=1.64$, $\mu_L^{Co}=0.1$, $\mu_S^{Fe}=2.17$ and $\mu_L^{Fe}=0.065$, in the case of Co/Fe bilayer, and $\mu_S^{Co2}=1.77$, $\mu_L^{Co2}=0.1$,  $\mu_S^{Co1}=1.58$, $\mu_L^{Co1}=0.066$, $\mu_S^{Fe}=2.3$ and $\mu_L^{Fe}=0.074$ in the case of Co$_2$/Co$_1$/Fe trilayer. All values are given in Bohr magneton $\mu_B$. The calculations also predict an in-plane magnetic anisotropy for all the studied structures. For instance, we obtain magnetic anisotropy energies of  $-0.69$ and $-1.77$, and $-0.77$~meV for the Co/Ni/Ir(001), Co/Co/Fe/Ir(001) and Co/Fe/Co/Ir(001) multilayer structures, respectively. The negative sign denotes an in-plane easy axis, in full agreement with our MOKE analysis.

\threesubsection{Magnon energy asymmetry}

In the experiment the magnon energy asymmetry $\Delta\varepsilon(Q)$ was measured. In order to have a quantitative analysis we calculated this quantity using the values of $\vec{D}_{ij}$s obtained from the first-principles calculations and compared the results to those of the experiment (see for example Figure~\ref{Fig3:DMIDispersion}).  Within a simple spin Hamiltonian model $\Delta\varepsilon(Q)$ is given by \cite{Zakeri2023a}.

\begin{eqnarray}
	\Delta\varepsilon(\vec{Q}) =  \sum_{\alpha}\vec{D}_{\alpha} \cdot \vec{\hat{e}} \sum_{\beta} \sin \left(\vec{Q} \cdot \vec{R}_{\beta}\right),
	\label{Eq:DispersionRelationDM}
\end{eqnarray}
where the unit vector $\vec{\hat{e}}$ represents the direction of the static magnetization, $\vec{D}_{\alpha}$ represents the DM vector of the $\alpha$'s   neighbor,  $\vec{R}_{\beta}$s represent the position vectors of all spins which are regarded as the $\alpha$'s neighbor (or shell). The first summation is over the number of neighbors and can be expanded over many different neighbors. The second summation is over the number of sites which belong to the same type of neighbor or shell.

\threesubsection{Statistical Analysis}

As described above, the magnon energies were extracted from the experimental difference spectra ($I_\downarrow-I_\uparrow$, see for example Fig.~\ref{Fig2:Difference}) by fitting them using a convolution of a Lorentzian and a Gaussian. The error bars in the magnons' energy are, therefore, given by the statistical and systematic uncertainties. They are given by the goodness of the fits and the deviations between the results of measurements repeated at a certain wavevector. The systematic error bars of $\varepsilon(Q)$ are estimated by the uncertainty in the magnon momentum, i.e., the momentum resolution. In our experiment the momentum resolution is $\Delta Q=0.02$ \AA$^{-1}$. This leads to a systematic error of a certain magnon with the energy of $\varepsilon(Q_0)$ and the wavevector $Q_0$ being $\frac{\partial\varepsilon(Q)}{\partial Q} |_{Q=Q_0}\Delta Q$, where $\frac{\partial\varepsilon(Q)}{\partial Q} |_{Q=Q_0}$ represents the slope of the magnon dispersion relation at $Q_0$ and $\varepsilon(Q_0)$. In order to eliminate such errors associated with the scattering geometry for probing magnons with the opposite $Q$ the sample magnetization to the opposite direction, keeping the scattering geometry unchanged.

The error bars in Fig.~\ref{Fig3:DMIDispersion} represent the uncertainties in the determination of $\Delta\varepsilon(Q)$. They include errors in the determination of the magnon energies for each magnetization direction.  Slightly different magnon energies were obtained when using different fitting protocols to evaluate the experimental data. In other words the error bars represent the uncertainty in the upper limit of the absolute values of $\Delta\varepsilon(Q)$. They are, in fact, largely overestimated. Our intention was to show the data as they are evaluated. Hence, no additional statistical corrections of the error bars were performed. Since each data point is measured  two or three times, the standard errors should, in principle, be by a factor of $\sqrt{2}$ or $\sqrt{3}$ smaller than the error bars shown in Fig.~\ref{Fig3:DMIDispersion}.  The density of the data points measured on several samples clearly shows the experimental trend and justifies the main message of the manuscript.

\medskip
\textbf{Supporting Information} \par 
Supporting Information is available from the Wiley Online Library or from the corresponding author.

\medskip
\textbf{Acknowledgements} \par 
This study was financially supported by the Deutsche Forschungsgemeinschaft (DFG) through the DFG Grants ZA 902/7-1, ZA 902/8-1 and ZA 902/9-1. Kh.Z. thanks the Physikalisches Institut for hosting the group and providing the necessary infrastructure. The authors thank Dr. Alberto Marmodoro for initiating the fruitful collaboration between the teams in Karlsruhe and Munich. The authors acknowledge support by the KIT-Publication Fund of the Karlsruhe Institute of Technology.

Open access funding enabled and organized by Projekt DEAL.

\medskip
\textbf{Conflict of Interest} \par

The authors declare no conflict of interest.

\medskip
\textbf{Data Availability Statement} \par

The data that support the ﬁndings of this study are available from the corresponding author upon reasonable request.

\medskip
\textbf{Keywords} \par

Magnetic interactions, Dzyaloshinskii--Moriya interaction, chirality, chiral spin textures, magnons

\medskip

%

\bibliographystyle {MSP}
\bibliography{Refs}

\begin{thebibliography}{10}
\providecommand{\url}[1]{\texttt{#1}}
\providecommand{\urlprefix}{URL }

\bibitem{Naaman2011}
\emph{Electronic and Magnetic Properties of Chiral Molecules and Supramolecular
  Architectures},
\newblock Springer Berlin Heidelberg, \textbf{2011}.

\bibitem{Dzyaloshinsky1958}
I.~Dzyaloshinsky,
\newblock A thermodynamic theory of weak ferromagnetism of antiferromagnetics,
\newblock \emph{Journal of Physics and Chemistry of Solids} \textbf{1958},
  \emph{4}, 4 241 .

\bibitem{Moriya1960}
T.~Moriya,
\newblock Anisotropic superexchange interaction and weak ferromagnetism,
\newblock \emph{Phys. Rev.} \textbf{1960}, \emph{120} 91.

\bibitem{Grigoriev2013}
S.~V. Grigoriev, N.~M. Potapova, S.-A. Siegfried, V.~A. Dyadkin, E.~V. Moskvin,
  V.~Dmitriev, D.~Menzel, C.~D. Dewhurst, D.~Chernyshov, R.~A. Sadykov, L.~N.
  Fomicheva, A.~V. Tsvyashchenko,
\newblock Chiral properties of structure and magnetism in
  {Mn$_{1-x}$Fe$_{x}$Ge} compounds: When the left and the right are fighting,
  who wins?,
\newblock \emph{Physical Review Letters} \textbf{2013}, \emph{110}, 20 207201.

\bibitem{Shibata2013}
K.~Shibata, X.~Z. Yu, T.~Hara, D.~Morikawa, N.~Kanazawa, K.~Kimoto,
  S.~Ishiwata, Y.~Matsui, Y.~Tokura,
\newblock Towards control of the size and helicity of skyrmions in helimagnetic
  alloys by spin-orbit coupling,
\newblock \emph{Nature Nanotechnology} \textbf{2013}, \emph{8}, 10 723.

\bibitem{Siegfried2015}
S.-A. Siegfried, E.~V. Altynbaev, N.~M. Chubova, V.~Dyadkin, D.~Chernyshov,
  E.~V. Moskvin, D.~Menzel, A.~Heinemann, A.~Schreyer, S.~V. Grigoriev,
\newblock Controlling the {Dzyaloshinskii-Moriya} interaction to alter the
  chiral link between structure and magnetism for {Fe$_{1-x}$Co$_{x}$Si},
\newblock \emph{Physical Review B} \textbf{2015}, \emph{91}, 18 184406.

\bibitem{Beutier2017}
G.~Beutier, S.~Collins, O.~Dimitrova, V.~Dmitrienko, M.~Katsnelson,
  Y.~Kvashnin, A.~Lichtenstein, V.~Mazurenko, A.~Nisbet, E.~Ovchinnikova,
  D.~Pincini,
\newblock Band filling control of the {Dzyaloshinskii-Moriya} interaction in
  weakly ferromagnetic insulators,
\newblock \emph{Physical Review Letters} \textbf{2017}, \emph{119}, 16 167201.

\bibitem{Abdizadeh2020}
S.~Abdizadeh, J.~Abouie, K.~Zakeri,
\newblock Dynamical switching of confined magnetic skyrmions under circular
  magnetic fields,
\newblock \emph{Physical Review B} \textbf{2020}, \emph{101}, 2 024409.

\bibitem{Fert1980}
A.~Fert, P.~M. Levy,
\newblock Role of anisotropic exchange interactions in determining the
  properties of spin-glasses,
\newblock \emph{Phys. Rev. Lett.} \textbf{1980}, \emph{44} 1538.

\bibitem{Crepieux1998}
A.~Crepieux, C.~Lacroix,
\newblock Dzyaloshinsky--{M}oriya interactions induced by symmetry breaking at
  a surface,
\newblock \emph{J. Magn. Magn. Mater.} \textbf{1998}, \emph{182}, 3 341 .

\bibitem{Roessler2006}
U.~K. R\"o{\ss}ler, A.~N. Bogdanov, C.~Pfleiderer,
\newblock Spontaneous skyrmion ground states in magnetic metals,
\newblock \emph{Nature} \textbf{2006}, \emph{442}, 7104 797.

\bibitem{Bode2007}
M.~Bode, M.~Heide, K.~von Bergmann, P.~Ferriani, S.~Heinze, G.~Bihlmayer,
  A.~Kubetzka, O.~Pietzsch, S.~Bl\"ugel, R.~Wiesendanger,
\newblock Chiral magnetic order at surfaces driven by inversion asymmetry,
\newblock \emph{Nature} \textbf{2007}, \emph{447}, 7141 190.

\bibitem{Vedmedenko2007}
E.~Y. Vedmedenko, L.~Udvardi, P.~Weinberger, R.~Wiesendanger,
\newblock Chiral magnetic ordering in two-dimensional ferromagnets with
  competing {D}zyaloshinsky-{M}oriya interactions,
\newblock \emph{Phys. Rev. B} \textbf{2007}, \emph{75} 104431.

\bibitem{Heinze2011}
S.~Heinze, K.~von Bergmann, M.~Menzel, J.~Brede, A.~Kubetzka, R.~Wiesendanger,
  G.~Bihlmayer, S.~Bl\"ugel,
\newblock Spontaneous atomic-scale magnetic skyrmion lattice in two dimensions,
\newblock \emph{Nature Physics} \textbf{2011}, \emph{7}, 9 713.

\bibitem{Fert2017}
A.~Fert, N.~Reyren, V.~Cros,
\newblock Magnetic skyrmions: advances in physics and potential applications,
\newblock \emph{Nature Reviews Materials} \textbf{2017}, \emph{2}, 7 17031.

\bibitem{Chen2020}
G.~Chen, A.~Mascaraque, H.~Jia, B.~Zimmermann, M.~Robertson, R.~L. Conte,
  M.~Hoffmann, M.~A. Gonzalez~Barrio, H.~Ding, R.~Wiesendanger, E.~G. Michel,
  S.~Bl\"ugel, A.~K. Schmid, K.~Liu,
\newblock Large {D}zyaloshinskii-{M}oriya interaction induced by chemisorbed
  oxygen on a ferromagnet surface,
\newblock \emph{Science Advances} \textbf{2020}, \emph{6}, 33 eaba4924.

\bibitem{Kim2019}
D.-H. Kim, M.~Haruta, H.-W. Ko, G.~Go, H.-J. Park, T.~Nishimura, D.-Y. Kim,
  T.~Okuno, Y.~Hirata, Y.~Futakawa, H.~Yoshikawa, W.~Ham, S.~Kim, H.~Kurata,
  A.~Tsukamoto, Y.~Shiota, T.~Moriyama, S.-B. Choe, K.-J. Lee, T.~Ono,
\newblock Bulk {D}zyaloshinskii-{M}oriya interaction in amorphous ferrimagnetic
  alloys,
\newblock \emph{Nature Materials} \textbf{2019}, \emph{18}, 7 685.

\bibitem{Zhang2022}
Q.~Zhang, J.~Liang, K.~Bi, L.~Zhao, H.~Bai, Q.~Cui, H.-A. Zhou, H.~Bai,
  H.~Feng, W.~Song, G.~Chai, O.~Gladii, H.~Schultheiss, T.~Zhu, J.~Zhang,
  Y.~Peng, H.~Yang, W.~Jiang,
\newblock Quantifying the {D}zyaloshinskii-{M}oriya interaction induced by the
  bulk magnetic asymmetry,
\newblock \emph{Physical Review Letters} \textbf{2022}, \emph{128}, 16 167202.

\bibitem{Liang2022}
J.~Liang, M.~Chshiev, A.~Fert, H.~Yang,
\newblock Gradient-induced {D}zyaloshinskii-{M}oriya interaction,
\newblock \emph{Nano Letters} \textbf{2022}, \emph{22}, 24 10128.

\bibitem{Davydenko2019}
A.~V. Davydenko, A.~G. Kozlov, A.~G. Kolesnikov, M.~E. Stebliy, G.~S. Suslin,
  Y.~E. Vekovshinin, A.~V. Sadovnikov, S.~A. Nikitov,
\newblock {D}zyaloshinskii-{M}oriya interaction in symmetric epitaxial
  {[Co/Pd(111)]$_N$} superlattices with different numbers of {Co/Pd} bilayers,
\newblock \emph{Physical Review B} \textbf{2019}, \emph{99}, 1 014433.

\bibitem{Zhang2021}
Y.~Zhang, J.~Liu, Y.~Dong, S.~Wu, J.~Zhang, J.~Wang, J.~Lu, A.~R\"uckriegel,
  H.~Wang, R.~Duine, H.~Yu, Z.~Luo, K.~Shen, J.~Zhang,
\newblock Strain-driven {D}zyaloshinskii-{M}oriya interaction for
  room-temperature magnetic skyrmions,
\newblock \emph{Physical Review Letters} \textbf{2021}, \emph{127}, 11 117204.

\bibitem{Yang2015}
H.~Yang, A.~Thiaville, S.~Rohart, A.~Fert, M.~Chshiev,
\newblock Anatomy of {D}zyaloshinskii-{M}oriya interaction at {Co/Pt}
  interfaces,
\newblock \emph{Physical Review Letters} \textbf{2015}, \emph{115}, 26 267210.

\bibitem{Yang2018}
H.~Yang, O.~Boulle, V.~Cros, A.~Fert, M.~Chshiev,
\newblock Controlling {D}zyaloshinskii-{M}oriya interaction via chirality
  dependent atomic-layer stacking, insulator capping and electric field,
\newblock \emph{Scientific Reports} \textbf{2018}, \emph{8}, 1.

\bibitem{Zakeri2023a}
K.~Zakeri, A.~Marmodoro, A.~von Faber, S.~Mankovsky, H.~Ebert,
\newblock Chirality-inverted {D}zyaloshinskii-{M}oriya interaction,
\newblock \emph{Physical Review B} \textbf{2023}, \emph{108}, 10 l100403.

\bibitem{Tsurkan2020}
S.~Tsurkan, K.~Zakeri,
\newblock Giant {Dzyaloshinskii-Moriya} interaction in epitaxial {Co/Fe}
  bilayers with {$C_{2v}$} symmetry,
\newblock \emph{Physical Review B} \textbf{2020}, \emph{102}, 6 060406.

\bibitem{Kuepferling2023}
M.~Kuepferling, A.~Casiraghi, G.~Soares, G.~Durin, F.~Garcia-Sanchez, L.~Chen,
  C.~Back, C.~Marrows, S.~Tacchi, G.~Carlotti,
\newblock Measuring interfacial dzyaloshinskii-moriya interaction in ultrathin
  magnetic films,
\newblock \emph{Reviews of Modern Physics} \textbf{2023}, \emph{95}, 1 015003.

\bibitem{Zakeri2010}
K.~Zakeri, Y.~Zhang, J.~Prokop, T.-H. Chuang, N.~Sakr, W.~X. Tang,
  J.~Kirschner,
\newblock Asymmetric spin-wave dispersion on {F}e(110): {D}irect evidence of
  the {D}zyaloshinskii-{M}oriya interaction,
\newblock \emph{Phys. Rev. Lett.} \textbf{2010}, \emph{104}, 13 137203.

\bibitem{Zakeri2012}
K.~Zakeri, Y.~Zhang, T.-H. Chuang, J.~Kirschner,
\newblock Magnon lifetimes on the {F}e(110) surface: The role of spin-orbit
  coupling,
\newblock \emph{Phys. Rev. Lett.} \textbf{2012}, \emph{108}, 19 197205.

\bibitem{Zakeri2023}
K.~Zakeri, D.~Rau, J.~Jandke, F.~Yang, W.~Wulfhekel, C.~Berthod,
\newblock Direct probing of a large spin-orbit coupling in the {FeSe}
  superconducting monolayer on {STO},
\newblock \emph{ACS Nano} \textbf{2023}, \emph{17}, 10 9575.

\bibitem{Cho2015}
J.~Cho, N.-H. Kim, S.~Lee, J.-S. Kim, R.~Lavrijsen, A.~Solignac, Y.~Yin, D.-S.
  Han, N.~J.~J. van Hoof, H.~J.~M. Swagten, B.~Koopmans, C.-Y. You,
\newblock Thickness dependence of the interfacial {D}zyaloshinskii-{M}oriya
  interaction in inversion symmetry broken systems,
\newblock \emph{Nature Communications} \textbf{2015}, \emph{6}, 1 7635.

\bibitem{Chaurasiya2016}
A.~K. Chaurasiya, C.~Banerjee, S.~Pan, S.~Sahoo, S.~Choudhury, J.~Sinha,
  A.~Barman,
\newblock Direct observation of interfacial {D}zyaloshinskii-{M}oriya
  interaction from asymmetric spin-wave propagation in
  ${W}/{C}o{F}e{B}/{S}i{O}_2$ heterostructures down to sub-nanometer
  {C}o{F}e{B} thickness,
\newblock \emph{Scientific Reports} \textbf{2016}, \emph{6}, 1 32592.

\bibitem{LoConte2017}
R.~Lo~Conte, G.~V. Karnad, E.~Martinez, K.~Lee, N.-H. Kim, D.-S. Han, J.-S.
  Kim, S.~Prenzel, T.~Schulz, C.-Y. You, H.~J.~M. Swagten, M.~Kl\"aui,
\newblock Ferromagnetic layer thickness dependence of the dzyaloshinskii-moriya
  interaction and spin-orbit torques in ${P}t/{C}o/{A}l{O}_x$,
\newblock \emph{AIP Advances} \textbf{2017}, \emph{7}, 6.

\bibitem{Belmeguenai2018}
M.~Belmeguenai, Y.~Roussigne, H.~Bouloussa, S.~Cherif, A.~Stashkevich,
  M.~Nasui, M.~Gabor, A.~Mora-Hernandez, B.~Nicholson, O.-O. Inyang,
  A.~Hindmarch, L.~Bouchenoire,
\newblock Thickness dependence of the dzyaloshinskii-moriya interaction in
  co2feal ultrathin films: Effects of annealing temperature and heavy-metal
  material,
\newblock \emph{Physical Review Applied} \textbf{2018}, \emph{9}, 4 044044.

\bibitem{Ourdani2021}
D.~Ourdani, Y.~Roussigne, S.~M. Cherif, M.~S. Gabor, M.~Belmeguenai,
\newblock Dependence of the interfacial dzyaloshinskii-moriya interaction,
  perpendicular magnetic anisotropy, and damping in co-based systems on the
  thickness of pt and ir layers,
\newblock \emph{Physical Review B} \textbf{2021}, \emph{104}, 10 104421.

\bibitem{Zakeri2013a}
K.~Zakeri, T.-H. Chuang, A.~Ernst, L.~Sandratskii, P.~Buczek, H.~Qin, Y.~Zhang,
  J.~Kirschner,
\newblock Direct probing of the exchange interaction at buried interfaces,
\newblock \emph{Nature Nanotechnology} \textbf{2013}, \emph{8}, 11 853.

\bibitem{Zakeri2013}
K.~Zakeri, J.~Kirschner,
\newblock \emph{Probing Magnons by Spin-Polarized Electrons}, volume 125 of
  \emph{Topics in Applied Physics Magnonics From Fundamentals to Applications},
  chapter~7, 84 -- 99,
\newblock Springer, Berlin, Heidelberg, \textbf{2013}.

\bibitem{Zakeri2013b}
K.~Zakeri, Y.~Zhang, J.~Kirschner,
\newblock Surface magnons probed by spin-polarized electron energy loss
  spectroscopy,
\newblock \emph{Journal of Electron Spectroscopy and Related Phenomena}
  \textbf{2013}, \emph{189} 157.

\bibitem{Zakeri2014}
K.~Zakeri,
\newblock Elementary spin excitations in ultrathin itinerant magnets,
\newblock \emph{Phys. Rep.} \textbf{2014}, \emph{545} 47.

\bibitem{Zhang2011}
Y.~Zhang, P.~A. Ignatiev, J.~Prokop, I.~Tudosa, T.~R.~F. Peixoto, W.~X. Tang,
  K.~Zakeri, V.~S. Stepanyuk, J.~Kirschner,
\newblock Elementary excitations at magnetic surfaces and their spin
  dependence.,
\newblock \emph{Phys. Rev. Lett.} \textbf{2011}, \emph{106}, 12 127201.

\bibitem{Zhang2012}
Y.~Zhang, T.-H. Chuang, K.~Zakeri, J.~Kirschner,
\newblock Relaxation time of terahertz magnons excited at ferromagnetic
  surfaces,
\newblock \emph{Phys. Rev. Lett.} \textbf{2012}, \emph{109}, 8 087203.

\bibitem{Zakeri2021a}
K.~Zakeri, A.~Hjelt, I.~Maznichenko, P.~Buczek, A.~Ernst,
\newblock Nonlinear decay of quantum confined magnons in itinerant
  ferromagnets,
\newblock \emph{Phys. Rev. Lett.} \textbf{2021}, \emph{126}, 17 177203.

\bibitem{Udvardi2009}
L.~Udvardi, L.~Szunyogh,
\newblock Chiral asymmetry of the spin-wave spectra in ultrathin magnetic
  films,
\newblock \emph{Phys. Rev. Lett.} \textbf{2009}, \emph{102} 207204.

\bibitem{Costa2010a}
A.~T. Costa, R.~B. Muniz, S.~Lounis, A.~B. Klautau, D.~L. Mills,
\newblock Spin-orbit coupling and spin waves in ultrathin ferromagnets: {T}he
  spin-wave {R}ashba effect,
\newblock \emph{Phys. Rev. B} \textbf{2010}, \emph{82} 014428.

\bibitem{Zakeri2017}
K.~Zakeri,
\newblock Probing of the interfacial {H}eisenberg and
  {D}zyaloshinskii--{M}oriya exchange interaction by magnon spectroscopy,
\newblock \emph{Journal of Physics: Condensed Matter} \textbf{2017}, \emph{29},
  1 013001.

\bibitem{Costa2020}
M.~Costa, N.~M.~R. Peres, J.~Fernandez-Rossier, A.~T. Costa,
\newblock Nonreciprocal magnons in a two-dimensional crystal with out-of-plane
  magnetization,
\newblock \emph{Physical Review B} \textbf{2020}, \emph{102}, 1 014450.

\bibitem{Chuang2012}
T.-H. Chuang, K.~Zakeri, A.~Ernst, L.~Sandratskii, P.~Buczek, Y.~Zhang, H.~Qin,
  W.~Adeagbo, W.~Hergert, J.~Kirschner,
\newblock Impact of atomic structure on the magnon dispersion relation: A
  comparison between {Fe(111)/Au/W(110)},
\newblock \emph{Phys. Rev. Lett.} \textbf{2012}, \emph{109}, 20 207201.

\bibitem{Zakeri2024a}
K.~Zakeri, A.~von Faber, A.~Ernst,
\newblock Magnons and fundamental magnetic interactions in a ferromagnetic
  monolayer: {T}he case of the {N}i monolayer,
\newblock \emph{Phys. Rev. B} \textbf{2024}, \emph{109} L180406.

\bibitem{Belabbes2016}
A.~Belabbes, G.~Bihlmayer, F.~Bechstedt, S.~Bl\"ugel, A.~Manchon,
\newblock Hund's rule-driven {D}zyaloshinskii-{M}oriya interaction at
  $3d$--$5d$ interfaces,
\newblock \emph{Physical Review Letters} \textbf{2016}, \emph{117}, 24 247202.

\bibitem{Zhu2022}
L.~Zhu, L.~Zhu, X.~Ma, X.~Li, R.~A. Buhrman,
\newblock Critical role of orbital hybridization in the
  {D}zyaloshinskii-{M}oriya interaction of magnetic interfaces,
\newblock \emph{Communications Physics} \textbf{2022}, \emph{5}, 1 151.

\bibitem{Chuang2014}
T.-H. Chuang, K.~Zakeri, A.~Ernst, Y.~Zhang, H.~J. Qin, Y.~Meng, Y.-J. Chen,
  J.~Kirschner,
\newblock Magnetic properties and magnon excitations in {Fe(001)} films grown
  on {Ir(001)},
\newblock \emph{Phys. Rev. B} \textbf{2014}, \emph{89}, 17 174404.

\bibitem{Kim2018}
S.~Kim, K.~Ueda, G.~Go, P.-H. Jang, K.-J. Lee, A.~Belabbes, A.~Manchon,
  M.~Suzuki, Y.~Kotani, T.~Nakamura, K.~Nakamura, T.~Koyama, D.~Chiba, K.~T.
  Yamada, D.-H. Kim, T.~Moriyama, K.-J. Kim, T.~Ono,
\newblock Correlation of the {D}zyaloshinskii{\textendash}{M}oriya interaction
  with {H}eisenberg exchange and orbital asphericity,
\newblock \emph{Nature Communications} \textbf{2018}, \emph{9}, 1 1648.

\bibitem{Zakeri2024}
K.~Zakeri, A.~von Faber,
\newblock Giant spin-orbit induced magnon nonreciprocity in ultrathin
  ferromagnets,
\newblock \emph{Phys. Rev. Lett.} \textbf{2024}, \emph{132}, 12 126702.

\bibitem{Faber2024}
A.~von Faber, C.~Hins, K.~Zakeri,
\newblock Ubiquity of the spin-orbit induced magnon nonreciprocity in ultrathin
  ferromagnets,
\newblock \emph{Physical Review B} \textbf{2024}, \emph{110}, 13 134417.

\bibitem{Herve2018}
M.~Herve, B.~Dupe, R.~Lopes, M.~Boettcher, M.~D. Martins, T.~Balashov,
  L.~Gerhard, J.~Sinova, W.~Wulfhekel,
\newblock Stabilizing spin spirals and isolated skyrmions at low magnetic field
  exploiting vanishing magnetic anisotropy,
\newblock \emph{Nature Communications} \textbf{2018}, \emph{9}, 1 1015.

\bibitem{Qin2017}
H.~J. Qin, K.~Zakeri, A.~Ernst, J.~Kirschner,
\newblock Temperature dependence of magnetic excitations: Terahertz magnons
  above the {C}urie temperature,
\newblock \emph{Physical Review Letters} \textbf{2017}, \emph{118}, 12 127203.

\bibitem{Zhang2010}
Y.~Zhang, P.~Buczek, L.~Sandratskii, W.~X. Tang, J.~Prokop, I.~Tudosa, T.~R.~F.
  Peixoto, K.~Zakeri, J.~Kirschner,
\newblock Nonmonotonic thickness dependence of spin wave energy in ultrathin fe
  films: Experiment and theory,
\newblock \emph{Phys. Rev. B} \textbf{2010}, \emph{81}, 9 094438.

\bibitem{Prokop2009}
J.~Prokop, W.~X. Tang, Y.~Zhang, I.~Tudosa, T.~R.~F. Peixoto, K.~Zakeri,
  J.~Kirschner,
\newblock Magnons in a ferromagnetic monolayer,
\newblock \emph{Phys. Rev. Lett.} \textbf{2009}, \emph{102}, 17 177206.

\bibitem{Zakeri2014a}
K.~Zakeri, J.~Prokop, Y.~Zhang, J.~Kirschner,
\newblock Magnetic excitations in ultrathin magnetic films: Temperature
  effects,
\newblock \emph{Surface Science} \textbf{2014}, \emph{630} 311.

\bibitem{Qin2015}
H.~J. Qin, K.~Zakeri, A.~Ernst, L.~M. Sandratskii, P.~Buczek, A.~Marmodoro,
  T.~H. Chuang, Y.~Zhang, J.~Kirschner,
\newblock Long-living terahertz magnons in ultrathin metallic ferromagnets,
\newblock \emph{Nat. Commun.} \textbf{2015}, \emph{6} 6126.

\bibitem{Zakeri2010a}
K.~Zakeri, T.~Peixoto, Y.~Zhang, J.~Prokop, J.~Kirschner,
\newblock On the preparation of clean tungsten single crystals,
\newblock \emph{Surface Science} \textbf{2010}, \emph{604} L1.

\bibitem{Chen2017}
Y.-J. Chen, K.~Zakeri, A.~Ernst, H.~J. Qin, Y.~Meng, J.~Kirschner,
\newblock Group velocity engineering of confined ultrafast magnons,
\newblock \emph{Phys. Rev. Lett.} \textbf{2017}, \emph{119}, 26 267201.

\bibitem{Zakeri2021}
K.~Zakeri, H.~Qin, A.~Ernst,
\newblock Unconventional magnonic surface and interface states in layered
  ferromagnets,
\newblock \emph{Commun. Phys.} \textbf{2021}, \emph{4}, 1 18.

\bibitem{Heinz2009}
K.~Heinz, L.~Hammer,
\newblock Nanostructure formation on {I}r(100),
\newblock \emph{Prog. Surf. Sci.} \textbf{2009}, \emph{84}, 1 2 .

\bibitem{Tian2009}
Z.~Tian, D.~Sander, J.~Kirschner,
\newblock Nonlinear magnetoelastic coupling of epitaxial layers of {F}e, {C}o,
  and {N}i on {I}r(100),
\newblock \emph{Physical Review B} \textbf{2009}, \emph{79}, 2 024432.

\bibitem{Kueppers1979}
J.~K\"uppers, H.~Michel,
\newblock Preparation of {I}r(100)--$1{\times}1$ surface structures by surface
  reactions and its reconstruction kinetics as determined with {LEED}, {UPS}
  and work function measurements,
\newblock \emph{Applications of Surface Science} \textbf{1979}, \emph{3}, 2
  179.

\bibitem{Zakeri2024c}
K.~Zakeri, A.~Ernst,
\newblock Generation and propagation of ultrafast terahertz magnons in
  atomically architectured nanomagnets,
\newblock \emph{Nano Letters} \textbf{2024}, \emph{24}, 31 9528.

\bibitem{Zakeri2018}
K.~Zakeri,
\newblock Terahertz magnonics: Feasibility of using terahertz magnons for
  information processing,
\newblock \emph{Physica C-superconductivity and Its Applications}
  \textbf{2018}, \emph{549} 164.

\bibitem{Vollmer2003}
R.~Vollmer, M.~Etzkorn, P.~S.~A. Kumar, H.~Ibach, J.~Kirschner,
\newblock Spin-polarized electron energy loss spectroscopy of high energy,
  large wave vector spin waves in ultrathin fcc {Co} films on {Cu(001)},
\newblock \emph{Phys. Rev. Lett.} \textbf{2003}, \emph{91} 147201.

\bibitem{Qin2019}
H.~J. Qin, S.~Tsurkan, A.~Ernst, K.~Zakeri,
\newblock Experimental realization of atomic-scale magnonic crystals,
\newblock \emph{Phys. Rev. Lett.} \textbf{2019}, \emph{123} 257202.

\bibitem{Zakeri2011}
K.~Zakeri, Y.~Zhang, J.~Prokop, T.~H. Chuang, W.~X. Tang, J.~Kirschner,
\newblock Magnon excitations in ultrathin {F}e layers: The influence of the
  {D}zyaloshinskii-{M}oriya interaction,
\newblock \emph{Journal of Physics: Conference Series} \textbf{2011},
  \emph{303} 012004.

\bibitem{Ebert2011}
H.~Ebert, D.~K\"odderitzsch, J.~Min{\'{a}}r,
\newblock Calculating condensed matter properties using the
  {KKR}-green{\textquotesingle}s function method{\textemdash}recent
  developments and applications,
\newblock \emph{Reports on Progress in Physics} \textbf{2011}, \emph{74}, 9
  096501.

\bibitem{Vosko1980}
S.~H. Vosko, L.~Wilk, M.~Nusair,
\newblock Accurate spin-dependent electron liquid correlation energies for
  local spin density calculations: a critical analysis,
\newblock \emph{Canadian Journal of Physics} \textbf{1980}, \emph{58}, 8 1200.

\bibitem{Liechtenstein1987}
A.~I. Liechtenstein, M.~I. Katsnelson, V.~P. Antropov, V.~A. Gubanov,
\newblock Local spin density functional approach to the theory of exchange
  interactions in ferromagnetic metals and alloys,
\newblock \emph{J. Magn. Magn. Mater.} \textbf{1987}, \emph{67} 65.

\bibitem{Mankovsky2017}
S.~Mankovsky, H.~Ebert,
\newblock Accurate scheme to calculate the interatomic
  {D}zyaloshinskii-{M}oriya interaction parameters,
\newblock \emph{Physical Review B} \textbf{2017}, \emph{96}, 10 104416.

\end{thebibliography}



\begin{figure}
\textbf{Table of Contents}\\
\medskip
\includegraphics{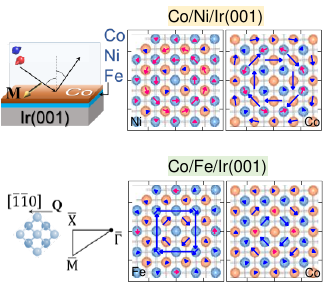}
\medskip
\caption*{In layered magnets, the atomistic Dzyaloshinskii–Moriya interaction (DMI) depends critically not only on the orbital occupancy of the interface layer but also on the sequence of the atomic layers. The effect can be understood by analyzing the contributions of different orbitals to DMI. Both the chirality and the magnitude of the atomistic DMI can be controlled through interface engineering.}
\end{figure}

\end{document}